\documentclass[12pt]{article}        
\usepackage{amstex}
\usepackage{amssymb}
\usepackage{cite}
\usepackage{epsfig}

\begin{document}                     

\allowdisplaybreaks

\begin{titlepage}


\begin{flushright}
UTHEP-97-0102\\
CERN-TH-97-48
\end{flushright}

\vspace{1mm}
\begin{center}
{\Large\bf
$e^+e^-$ Annihilation into  Hadrons at LEP2\\
 in the Presence of
 the Anomalous DESY Positron-Jet Event Phenomenon
}
\end{center}
\vspace{3mm}

\begin{center}
{\bf S. Jadach$^{a,b}$},
{\bf B.F.L. Ward$^{a,c,d}$}
{\em and}
{\bf Z. W\c as$^{a,b}$} \\

\vspace{1mm}
{\em $^a$CERN, Theory Division, CH-1211 Geneva 23, Switzerland,}\\
{\em $^b$Institute of Nuclear Physics,
  ul. Kawiory 26a, Krak\'ow, Poland,}\\
{\em $^c$Department of Physics and Astronomy,\\
  The University of Tennessee, Knoxville, Tennessee 37996-1200,\\
  $^d$SLAC, Stanford University, Stanford, California 94309} 
\end{center}

\vspace{10mm}
\begin{abstract}
We discuss the allowed parameter regime in the coupling-mass
plane implied by the existing LEP2 data on 
$e^+e^-\rightarrow hadrons$ at 172.3 GeV
for the leptoquark interpretation of the
anomalous DESY positron-jet events for four different models
of the leptoquark charges and chiral couplings to quarks,
for both a loose cut on $s'/s$ and a tight cut
on $s'/s$. We find that this interpretation of the DESY
phenomenon is still consistent with the LEP2 data although,
for vector leptoquarks, 
a significant regime of the relevant parameter space is already excluded.
\end{abstract}
 
\begin{center}
{\it Submitted to Phys. Lett. B}
\end{center}

\vspace{1mm}
\begin{flushleft}
{\bf CERN-TH/97-48\\
     March,~1997}
\end{flushleft}

\renewcommand{\baselinestretch}{0.1}
\footnoterule
\noindent
{\footnotesize
\begin{itemize}
\item[${\dagger}$]
Work supported in part by 
the US DoE contract DE-FG05-91ER40627 and DE-AC03-76SF00515,
Polish Government grants KBN 2P03B17210,
the European Union under contract No. ERB-CIPD-CT94-0016,
and Polish-French Collaboration within IN2P3.
\end{itemize}
}
 
 
\end{titlepage}

Recently, the H1 and ZEUS Collaborations have reported~\cite{h1,zeus}
an anomalous amount of high $Q^2$, high $x_{Bj}$ events in the
deep inelastic $e^+$p scattering at HERA. In this paper, we investigate
the consequences in the LEP2 process
$e^+e^-\rightarrow hadrons$ of the existence of the DESY
phenomenon if it is interpreted as the exchange of a spin one
or zero object
in the s-channel in the reduced 
$e^+{}^{\tiny(}\bar q^{\tiny)}$
parton level process at HERA. For, by the crossing symmetry,
this object, whatever it may be, should then be exchanged in the t(u)-channel
in the LEP2 $e^+e^-$ annihilation and the current agreement level between
the SM (Standard Model) 
and the LEP2 data on $e^+e^-\rightarrow hadrons$ should already
place some constraints on the allowed couplings and mass ranges.
In our work, we elucidate some of these constraints. We note that
preliminary constraints on a scalar leptoquark from
LEP2 $e^+e^-$ annihilation data have already been presented
by S. Kamamiya in Ref.~\cite{jambore97}. We will see that
our analysis of the vector and scalar cases 
yields results consistent with the latter.

For definiteness, we shall call this spin one or zero object a `leptoquark',
since according to Ref.~\cite{H1-sem}, it has a mass $\sim 200 GeV$
and a width $\lesssim 2.2$ GeV. We stress as it has been already done in
Ref.~\cite{buchmuller:1987} that as long as the couplings of this
object, which may be composite or elementary, are sufficiently
chiral and diagonal in flavor and zero on diquark fields, 
it is not in contradiction with any known physical requirement.
Thus, we proceed entirely phenomenologically and try to answer the
very definite question as to whether the LEP2 data are consistent
with a vector or scalar `leptoquark' explanation of the DESY data.
See Refs.~\cite{guido:1997,pete:1997,dreiner:1997,godfrey:1997,blumlein:1997}
for related but independent analyses of the DESY data and its possible
interpretation. We recall that the scalar leptoquark and the
susy scalar quark are, from the point of view of our analysis, essentially
synonymous~\cite{guido:1997,pete:1997,dreiner:1997}.

More specifically, in this brief note we record the 
differential cross section for
$e^+e^-\rightarrow \bar q q$ in the presence of the DESY leptoquark
for two models of leptoquark charges. We have, for the quark
elemental solid angle $d\Omega_q$ in the cms system,
\begin{equation}
{d\sigma\over d\Omega_q}= {1\over 64\pi^2s}|\bar {\cal M}|^2,
\label{sig}
\end{equation}
where the spin averaged squared matrix element, 
for the leptoquark charges $({5\over 3},{2\over 3})$ 
coupling to left-handed quarks (model (1)), is
\begin{equation}
\label{matsq1L}
\begin{split}
|\bar {\cal M}|^2_{ \{ {5\over 3}, {2\over3}  \}_L }
&= \left| {\delta_{S,1}g_X^2\over 2D_X(t)}+\sum_{\rho=A,Z} {G_\rho^2
    (v_\rho(q)+a_\rho(q))(v_\rho(e)+a_\rho(e))\over D_\rho(s)}\right|^2 u^2
\\
&+ \left| \sum_{\rho=A,Z} {G_\rho^2
    (v_\rho(q)-a_\rho(q))(v_\rho(e)-a_\rho(e))\over D_\rho(s)}\right|^2 u^2
\\&
+ \left| \sum_{\rho=A,Z} {G_\rho^2
    (v_\rho(q)-a_\rho(q))(v_\rho(e)+a_\rho(e))\over D_\rho(s)}\right|^2 t^2
\\
&+ \left| {\delta_{S,0}g_X^2\over 4D_X(t)}+\sum_{\rho=A,Z} {G_\rho^2
    (v_\rho(q)+a_\rho(q))(v_\rho(e)-a_\rho(e))\over D_\rho(s)}\right|^2 t^2
,
\end{split}
\end{equation}
for the leptoquark charges $({5\over 3},{2\over 3})$ 
coupling to right-handed quarks (model (2)), is
\begin{equation}
\label{matsq1R}
\begin{split}
|\bar {\cal M}|^2_{ \{ {5\over 3}, {2\over3}  \}_R }
&= \left| \sum_{\rho=A,Z} {G_\rho^2
    (v_\rho(q)+a_\rho(q))(v_\rho(e)+a_\rho(e))\over D_\rho(s)}\right|^2 u^2
\\
&+ \left| {\delta_{S,1}g_X^2\over 2D_X(t)}+\sum_{\rho=A,Z} {G_\rho^2
    (v_\rho(q)-a_\rho(q))(v_\rho(e)-a_\rho(e))\over D_\rho(s)}\right|^2 u^2
\\
&+ \left| {\delta_{S,0}g_X^2\over 4D_X(t)}+\sum_{\rho=A,Z} {G_\rho^2
    (v_\rho(q)-a_\rho(q))(v_\rho(e)+a_\rho(e))\over D_\rho(s)}\right|^2 t^2
\\
&+ \left| \sum_{\rho=A,Z} {G_\rho^2
    (v_\rho(q)+a_\rho(q))(v_\rho(e)-a_\rho(e))\over D_\rho(s)}\right|^2 t^2
,
\end{split}
\end{equation}
and, for the leptoquark charges $({4\over 3},{1\over 3})$  
coupling to left-handed quarks (model (3)), is
\begin{equation}
\label{matsq2L}
\begin{split}
|\bar {\cal M}|^2_{ \{ {4\over 3}, {1\over3}  \}_L }
 &= \left| \sum_{\rho=A,Z} {G_\rho^2
    (v_\rho(q)-a_\rho(q))(v_\rho(e)+a_\rho(e))\over D_\rho(s)}\right|^2 t^2
\\
&+ \left| -{\delta_{S,1}g_X^2\over 2D_X(u)}+\sum_{\rho=A,Z} {G_\rho^2
    (v_\rho(q)+a_\rho(q))(v_\rho(e)-a_\rho(e))\over D_\rho(s)}\right|^2 t^2
\\
&+ \left| -{\delta_{S,0}g_X^2\over 4D_X(u)}+\sum_{\rho=A,Z} {G_\rho^2
    (v_\rho(q)+a_\rho(q))(v_\rho(e)+a_\rho(e))\over D_\rho(s)}\right|^2 u^2
\\
&+ \left| \sum_{\rho=A,Z} {G_\rho^2
    (v_\rho(q)-a_\rho(q))(v_\rho(e)-a_\rho(e))\over D_\rho(s)}\right|^2 u^2
,
\end{split}
\end{equation}
and, for the leptoquark charges $({4\over 3},{1\over 3})$  
coupling to right-handed quarks (model (4)), is
\begin{equation}
\label{matsq2R}
\begin{split}
|\bar {\cal M}|^2_{ \{ {4\over 3}, {1\over3}  \}_R }
 &= \left| -{\delta_{S,1}g_X^2\over 2D_X(u)}+\sum_{\rho=A,Z} {G_\rho^2
    (v_\rho(q)-a_\rho(q))(v_\rho(e)+a_\rho(e))\over D_\rho(s)}\right|^2 t^2
\\
&+ \left| \sum_{\rho=A,Z} {G_\rho^2
    (v_\rho(q)+a_\rho(q))(v_\rho(e)-a_\rho(e))\over D_\rho(s)}\right|^2 t^2
\\
&+ \left| \sum_{\rho=A,Z} {G_\rho^2
    (v_\rho(q)+a_\rho(q))(v_\rho(e)+a_\rho(e))\over D_\rho(s)}\right|^2 u^2
\\
&+ \left| -{\delta_{S,0}g_X^2\over 4D_X(u)}+\sum_{\rho=A,Z} {G_\rho^2
    (v_\rho(q)-a_\rho(q))(v_\rho(e)-a_\rho(e))\over D_\rho(s)}\right|^2 u^2
,
\end{split}
\end{equation}
where $S=0,1$ is the spin of the leptoquark $X$ and $\delta_{a,b}$
is the Kronecker delta function and where
we have defined the following kinematical and dynamical variables:
\begin{equation}
\label{parm}
\begin{split}
& D_X(u) = u-M_X^2, \quad M_X= 200~GeV,\\
& D_Z(s) = s-M_Z^2+i\Gamma_Zs/M_Z,\quad
  D_A(s) = s,\\
& v_Z(f) = {1\over2}I_3-Q_f\sin^2\theta_W,\quad
  a_Z(f) = {1\over2}I_3,\\
&  v_A(f) =Q_f,\quad
   a_A(f) =0,\\
& G_Z = e/(\sin\theta_W\cos\theta_W),\quad
  G_A = e,\\
& g_X = (e/\sin\theta_W)(1 + \delta),\\
&   s = (p_1+q_1)^2, ~t~=~(p_1-p_2)^2, ~u~=~(p_1-q_2)^2,
\end{split}
\end{equation}
where $\{p_1,q_1\}$ are the incoming $e^+,e^-$ 4-momenta respectively
and $\{p_2,q_2\}$ are the outgoing $\{\bar q, q\}$ 4-momenta respectively.
Here, $\delta$ is unknown and is to be varied to see what the
data will allow. 
$I_3$ is the usual weak isospin 3-component for fermion f
and $Q_f$ is its electric charge in units of the positron charge $e$.
We have thus complied with the constraint 
from Refs.~\cite{buchmuller:1987,leurer:1994a,leurer:1994b} that
only quarks of a specific chirality should couple to any
particular leptoquark.

\begin{figure*}[!ht]
\centering
\setlength{\unitlength}{0.100mm}
\begin{picture}(1600,1600)
\put(0,0){\makebox(0,0)[lb]{
\epsfig{file=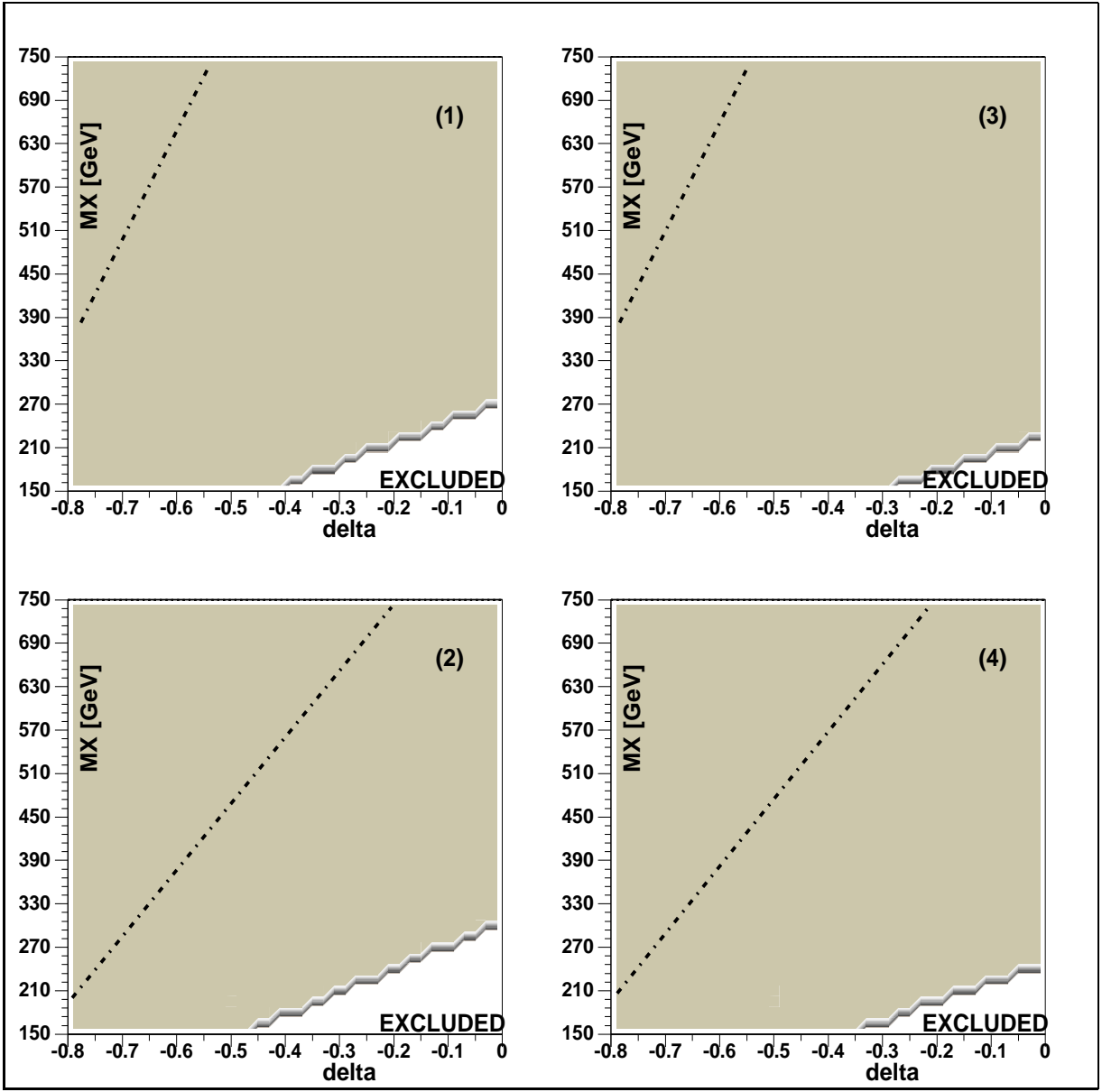,width=16.0cm,height=16cm}
}}
\put(0,0){\framebox( 1610,1600){ }}
\end{picture}
\vspace{-2mm}
\caption{\small\sf
Exclusion plots at \protect$95\%$ CL for the hard cut 
\protect$(s'/s)_{\min} \in (0.72,0.81)$
implied by the LEP2 172.3 GeV data on \protect$e^+e^-\rightarrow hadrons$
for the four vector leptoquark models described in the text: in
plot (i) we give the excluded region of the \protect$g_X-M_X$ plane
for model (i) as described in the text, $i=1,\cdots,4$,
with $g_X$ given by $\delta$ via (\protect\ref{parm}) and $M_X$ in GeV.
Dashed (dotted) line corresponds to limits of 
Ref.~\protect\cite{leurer:1994a} for spin 1
as explained in the text.
}
\label{fig:hardcut}
\end{figure*}

The formulae presented above we have implemented into 
the KORALZ Monte Carlo~\cite{koralz:1994} program.
We have performed the technical tests of the matrix elements as implemented
in KORALZ and compared them with analytical results on several approximated
forms of the above matrix elements,  e.g. for $Z+\gamma$ or only 
$Z$ exchange excluded,  
and for the production angle $\theta$ such as 
$\cos\theta=0, \pm{1 \over 2}$.
Agreement of 4-5 digits was always found.  
Later, QED and electroweak corrections were extended (to our 
X-exchange i.e. non s-channel Z, $\gamma$
interaction) automatically,
accordingly to the prescription identical to that, explained in
Ref.~\cite{colas:1990}. In particular the discussion of uncertainties 
of QED corrections implementation presented there
can be straightforwardly extended to the present case. 

We stress that
the leptoquark-quark-lepton vertices which we have assumed in 
eqs.~(\ref{matsq1L}-\ref{matsq2R}), corresponding to the $U_1,\tilde U_1$,
$\tilde V_2,V_2,$ and $U_3$ spin 1 and $R_2$, $\tilde R_2$, $S_1$, 
$\tilde S_1$ and $S_3$ spin 0
examples in Ref.~\cite{buchmuller:1987}, 
are intended to be generic
and not exhaustive: it is straightforward to include more general
coupling scenarios into our KORALZ~\cite{koralz:1994} 
calculational framework, should this
become necessary. We point-out further that
we may identify our spin 1(0) states in models 
(1-4) respectively with the corresponding charge and mass
eigenstates formed from the 
states $S_L$,$T$,
$\tilde S_R$, $S_R$, $D_L$,$D_R$ and $\tilde D$ in the 
notation of Refs.~\cite{leurer:1994a,leurer:1994b}; 
for, at scales $\sim 200$ GeV,
we expect the $SU_{2L}\times U_1$ electroweak symmetry to be broken with
leptoquark states of the same charge and color mixed into the respective
mass eigenstates and it is these mass eigenstates that we have
used in (\ref{sig}). In other words, from the DESY data we
we may have that $\bar e q$ resonates into the leptoquark $X,~q=u,d$,
models (1) and (2) with t-channel X exchange in $e^+e^-\rightarrow \bar{q}q$
and $F=0$ in the language of 
Refs.~\cite{buchmuller:1987,leurer:1994a,leurer:1994b},
or that $\bar e \bar q$ resonates 
into the leptoquark $X,~q=u,d$, models (3) and
(4) with u-channel X exchange in $e^+e^-\rightarrow \bar{q}q$ and
$F=-2$ in the language of 
Refs.~\cite{buchmuller:1987,leurer:1994a,leurer:1994b},
where F is the fermion number of X. 
For definiteness, we have assumed strong isospin symmetry for
simplicity; it is trivial to relax this last assumption, should more data
render this necessary.
For completeness, we then record the interaction Lagrangian densities
which we used to arrive at the results (\ref{matsq1L})--(\ref{matsq2R}), for
$S=1$,
\begin{equation}
\label{int}
\begin{split}
{\cal L}_{int,1,H}^{\{5/3,2/3\}} 
  &= -{g_X\over \sqrt 2}\left(X^{(-5/3)\alpha}_\mu
\bar{e}_H\gamma^\mu{u^\alpha_H}+ X^{(-2/3)\alpha}_\mu
\bar{e}_H\gamma^\mu{d^\alpha_H} + h.c.\right), 
\\
{\cal L}_{int,1,H}^{\{4/3,1/3\}} 
  &= -{g_X\over \sqrt 2}\left(X^{(-4/3)\alpha}_\mu
\bar{d}^\alpha_H\gamma^\mu{e^c_{H'}}+ X^{(-1/3)\alpha}_\mu
\bar{u}^\alpha_H\gamma^\mu{e^c_{H'}} + h.c.\right),
\end{split}
\end{equation}
where here $H=L,R$,
and $H'$ denotes chirality opposite to $H$, 
i.e. $H'=L$, for $H=R$ and $H'=R$ for $H=L$.
We have defined $\psi_{L(R)}=P_{L(R)}\psi$  as well as 
$\psi^c=C\bar\psi^T$ for all $\psi$, where $T$ denotes transposition. 
The the chiral projections are here $P_L=(1-\gamma^5)/2$,$P_R=(1+\gamma^5)/2$
and $C$ is the charge conjugation matrix in an appropriate representation.
Thus, ${\cal L}^{\{5/3,2/3\}}_{int,1,L}$ describes the interactions
used to derive the result (\ref{matsq1L}),~...~, and so on. The 
analogous formulas for $S=0$ for ${\cal L}^{\{Q,Q-1\}}_{int,0,H}$
follow from those for ${\cal L}^{\{Q,Q-1\}}_{int,1,H}$ from the
corresponding replacements $vector~X \times ~ vector~ Dirac~ current
\Leftrightarrow~ scalar~ X \times~ scalar~ Dirac~ current$ in an 
obvious way so we do not list them explicitly.

In the Fig.~1, we exhibit the 95\% CL exclusion plot
in the $g_X,M_X$ plane, for $S=1$, which follows from comparing the
cross sections in eqs.~(\ref{sig})--(\ref{matsq2R}) as implemented
in KORALZ~\cite{koralz:1994}
with the allowed deviation from the SM expectation,
where we take the respective leptoquark to couple to the first
generation only and we use the LEP2 data presented at the LEP
Jamboree~\cite{jambore97} to derive the allowed deviation 
$-0.0251\sigma_{SM}\le \Delta\sigma \le 0.119\sigma_{SM}$
for the tight cut%
\footnote{ In the process of combining experimental data
  from different LEP collaborations we have corrected data for slight
  difference in $(s'/s)_{\min}$.}
$(s'/s)_{\min} \in (0.72,0.81)$ and the allowed 
deviation $-0.0103\sigma_{SM}\le \Delta\sigma \le 0.0613\sigma_{SM}$
for the loose cut $(s'/s)_{\min} \in (0.010,0.015)$, both at 95\% CL.
Here, $\sigma_{SM}$ is the respective SM cross section
and $\Delta\sigma$ is the corresponding deviation allowed experimentally%
\footnote{ At $\protect\sqrt{s}=172.3$ GeV
  for $\protect\sqrt{(s'/s)_{\min}} = 0.85$
  we find from KORALZ $\sigma_{SM}= 28.43$pb
  and combined results of four LEP2 experiments~\protect\cite{jambore97} 
  (all errors combined in quadrature)
  give us $(\sigma-\sigma_{SM})/\sigma_{SM}=  0.0472 \pm 0.037$.
}.
For the latter excluded region, we do not find
any significant change over the limits that follow from the
former excluded region, so we only consider the former
one henceforth. Thus, for each of our
four leptoquark models, charges (5/3,2/3) with left-handed (model~1)
and right-handed (model~2) couplings to quarks and charges (4/3,1/3)
with left-handed (model~3) and right-handed (model 4) couplings
to quarks, we show the exclusion plots corresponding to the former
(hard $s'/s$ cut)
95\% CL deviation interval just given, for $S=1$.
These are shown in Fig.~1, for the $M_X$ range $[150~GeV,750~GeV]$ 
and for the coupling range corresponding to $-0.8\le \delta \le 0.0$ 
as defined in (\ref{parm}); for $S=0$, we do
find an excluded region in the parameter space shown in Fig.~1
from the present LEP2 data.  
We see that in all cases, already,
the LEP2 data rule out a significant part of the $g_X-M_X$ plane
for $S=1$
but that the $S=1$ leptoquark interpretation, like the $S=0$ case,
is still viable.
According to the data from DESY, one should focus on the region
below $M_X\cong 270$ GeV in Fig.~1. In this region, we see that
the size of the excluded parameter space depends on the model,
varying as it does, for spin 1 from a maximum of $\sim 34\%$ for model (2)
down to a minimum of $\sim 11\%$ for model (3); for spin 0,
the entire parameter space in Fig.~1 is allowed by present
LEP2 data; in particular, all of the mass region suggested by the
HERA data is allowed for $S=0$ by the present LEP2 data. 
For $\delta \lesssim -0.5$, the entire mass region
suggested by the DESY data is also still viable for $S=1$. For $\delta =0$,
the value of $M_X$ must be at least 220~GeV to be consistent with
our LEP2 data constraint for spin 1. 

The above results are consistent with those
presented in Ref.~\cite{jambore97} by Kamamiya for a scalar
leptoquark when due account is taken of our strong isospin
symmetry assumption and of our normalization of the respective
leptoquark coupling constant.
In Refs.~\cite{leurer:1994a,leurer:1994b}, indirect bounds on
leptoquarks were derived assuming only one such weak isospin
particle multiplet with a given chirality of its quark coupling
is present at a time. As we stated, in our work we do not
assume weak isospin symmetry. If we want to use the results in 
Refs.~\cite{leurer:1994a,leurer:1994b}, 
we need to make some assumption about the
leptoquark weak isospin mixing matrix in general. If we make
the simplest possible assumption, that is that our states
are composed of only those states in Refs.~\cite{leurer:1994a,leurer:1994b}, 
then for spin 1 we may identify 
$\{T^{(-5/3)},(T^{(-2/3)}-S^{-(2/3)}_L)/\sqrt{2}\}
     \Leftrightarrow \{X^{(-5/3)},X^{(-2/3)}\}$ 
in model (1),
$\{\tilde S^{(-5/3)},S^{(-2/3)}_R\}
     \Leftrightarrow \{X^{(-5/3)},X^{(-2/3)}\}$ 
in model (2),
$\{D^{(-4/3)}_L,D^{(-1/3)}_L\}
     \Leftrightarrow \{X^{(-4/3)},X^{(-1/3)}\}$ 
in model (3), and
$\{D^{(-4/3)}_R,\tilde D^{(-1/3)}\}
     \Leftrightarrow \{X^{(-4/3)},X^{(-1/3)}\}$ 
in model (4); for spin 0, an analogous transformation would hold.
These indentifications, with the attendant coupling constant relations
$2g=g_X,~\sqrt{2}g=g_X,~\sqrt{2}g=g_X$, and $\sqrt{2}g=g_X$, respectively
where $g$ is the coupling constant in Refs.~\cite{leurer:1994a,leurer:1994b},
lead to the bounds, from the formulas in 
Ref.~\cite{leurer:1994a,leurer:1994b}, for spin 1, 
    $M_X > (1+\delta)\; 1792$~GeV, 
    $M_X > (1+\delta)\;  956$~GeV,
    $M_X > (1+\delta)\; 1895$~GeV, 
and $M_X > (1+\delta)\;  956$~GeV, respectively,
for models (1-4); for spin 0, the corresponding results are
    $M_X > (1+\delta)\; 690$~GeV, 
    $M_X > (1+\delta)\; 1322$~GeV,
    $M_X > (1+\delta)\; 675$~GeV, 
and $M_X > (1+\delta)\; 1322$~GeV, respectively. 
We see that even with this simple mixing
assumption, the leptoquark interpretation is still viable
for $\delta \lesssim -0.75~(-0.55)$ for spin 1(0).
This result is extremely dependent
on the naive mixing assumption and on presuming only one model is present
at a time. For example, if we have both model (1) and model (2) present
for spin 1,
our bounds in Fig. ~1 would get stronger along the mass axis
by a factor $\sim \sqrt{2}$ but the Leurer bounds we just
derived would be obviated and replaced with much weaker bounds which
follow from the formulas leading to 
Tables 3 and 4 in Ref.~\cite{leurer:1994a}; 
in our models, the formula leading to 
the Table 2 in Ref.~\cite{leurer:1994a} does not
apply because we always have the relation $ \eta_{S_L}=\eta_T $,
in the notation of this last reference. 
We thus eagerly await more precise LEP2 data and more plentiful
DESY data in order to proceed further with our consistency check
between the two data sets. 
At this time, we see that the DESY
phenomenon is not inconsistent with the current level of agreement
between $\sigma_{SM}$ and $\sigma_{obs}$ for $e^+e^-\rightarrow hadrons$
at LEP2.
\par

It is important to view the results presented in Fig.~1 and in the
text above from the point of their relation to the data observed
at HERA. Indeed, concerning the $S=1$ case, it has been 
noted in Ref.~\cite{guido:1997} that Tevatron data ( see the
Note Added below) require that the branching 
ratio $B(X\rightarrow
e^+{}^(\bar q^))\equiv B(e^+{}^(\bar q^))$ 
should be much less than 1 for $M_X\simeq 200$GeV
unless the small signal has been missed for some reason~\cite{blumlein:1997}.
We therefore need to comment on the interplay of our excluded
region and the size of $B(e^+{}^(\bar q^))$ in relation to the
actual size of the apparent signal at HERA. In a recent analysis,
two of us (S.J. and B.F.L.W.) together with W. Placzek~\cite{sjbwwp:1997}
have analyzed the HERA data using our YFS exponentiated MC
event generator methods~\cite{sjw:1988} as they are 
realized in the event generator
LESKO-YFS~\cite{lesko-yfs:1992,zeuthen:1992,placzekPHD:1993}. 
In Ref.~\cite{sjbwwp:1997}
it is shown that, for $\Gamma_X\simeq 2$GeV, the value of $\delta$ which
reproduces the HERA signal best is $\sim -.72(-.68)$ for $S=0(1)$ respectively,
corresponding to the $B(e^+ q)\cong 1.1\%(.9\%)$, thereby
obviating the Tevatron $S=1$ bounds within the allowed regions presented
above. For $B(e^+ q)\cong 1$, the value of $\delta$ which
reproduces the HERA signal best~\cite{sjbwwp:1997} is $\sim -.95$ for $S=0,1$ 
, corresponding to the expected much smaller coupling
$g_X\sim .05g_W$ in comparison to the optimal coupling
in the small $B(e^+ q)$, $\Gamma_X\cong 2$GeV case, 
where $g_X\cong .3g_W$. The $S=1$ unit $e^+q$ branching ratio
case, while still allowed by our results above, is, according to
Ref.~\cite{guido:1997}, already excluded by Tevatron data.\par

As a final point, we point-out that
for $\delta=-0.9$ we do not observe sensitivity
of our observable to the leptoquark effect, even for $M_X=100$ GeV. 
Note that in this case the overall
normalization of the leptoquark effect in the amplitudes is 
reduced by factor of 100; thus, such a
loss of sensitivity should be expected.
Indeed, for negative $\delta$ approaching $-1$ simple  scaling of 
the mass limit
is not to be expected.
The effect of the $X$(leptoquark) amplitude becomes smaller in 
a non-linear
way. It becomes more and more profound in respectively
forward/backward directions. Effects of beam pipe cuts as well as effects
on angular distributions then become essential in establishing
data sensitivity in this case. The implied
kind of the extended study is rather easy to perform as we have a full 
Monte Carlo at our disposal,  but it is definitely beyond this quick note.  
It should be noted that, in many cases, 
e.g.  \cite{boillot:1989,dittmar:1997},
it was shown that such angular effects 
improve significance substantially. We hope to participate
in such studies as well elsewhere~\cite{elsewh:1997}.

In summary, we have investigated the constraints placed by LEP2 data
on the leptoquark interpretation of the anomalous positron-jet
phenomenon at DESY. We used our KORALZ~\cite{koralz:1994} 
Monte Carlo event generator
so that higher order radiative corrections to the Born level leptoquark
signal are calculated at the YFS exponentiated LL ${\cal O}(\alpha^2)$
$\bar\beta_0$ level, for both initial and final state radiation,
in the framework of Ref.~\cite{yfs3:1992}.
We find that, while a significant part of the spin 1
coupling-mass plane for the leptoquark is excluded, there still 
remains a large region of the plane that is viable for spin 1 and
all of the respective parameter space considered is allowed for spin 0:
$-1.0\le \delta \le 0.0$, $150GeV\le M_X$. We look forward
to more precise data which will address these remaining allowed regions.

\vspace{2mm}
\noindent
{\em Note Added:} As we were writing this paper, we became aware of the
work of J.~Kalinowski {\it et al.}~\cite{pete:1997}
in which the idea of a leptoquark interpretation of the DESY
anomalous positron-jet events is analyzed for its implications
in $e^+e^-\rightarrow hadrons$ at LEP2 energies; our work differs
from theirs in that we actually work with the realistic YFS
exponentiated LL ${\cal O}(\alpha^2)$ multiple photon radiatively
corrected predictions of the respective effects at LEP2 and
use them together with the available data to set exclusion
limits of the would-be leptoquark like object whereas their work
is at the Born level and gives generic expectations for the
respective effects in $e^+e^-\rightarrow hadrons$. Where the two
analyses overlap, they agree completely. We also recently
became aware of the work of G. Altarelli {\it et al.}~\cite{guido:1997}
and of J. Blumlein~\cite{blumlein:1997} in which the leptoquark
interpretation is discussed with attention to the constraints following
from TEVATRON data. The first of these latter works points-out that
such data apparently already exclude the vector leptoquark with the
mass $\sim 200$ GeV if it has either minimal vector or Yang-Mills type
couplings to gluons and an appreciable BR to the $e^+q$, respectively
$e^+\bar q$, final state; the second of these latter works argues
that the implied signal, depending as it does in detail on cuts, BR's, etc.,
may very well be missed and hence that
even under the stated assumptions the issue is still unsettled.
If these assumptions are not valid, 
the vector leptoquark interpretation would then be unequivocally viable at 
this time. Finally, we recently became aware of an
independent similar analysis by
M. Doncheski and S. Godfrey~\cite{godfrey:1997} who reach conclusions
similar to those we presented herein.

\vspace{2mm}
{\bf \Large Acknowledgements}\\
We are indebted to Prof. M. Veltman for initial discussions 
which led to the work presented in this paper.
Two of us (S.J. and B.F.L.W.) acknowledge 
the kind hospitality of Prof. G. Veneziano
and the CERN TH Division and the support of 
Prof. D. Schlatter and the ALEPH
Collaboration while this work was done.
One of us (Z. W.) acknowledges 
the kind hospitality and support, 
during the time while this work was done,
of the ALEPH Collaboration Group in LAL ORSAY.
We would like to thank Dr. W. P\l{}aczek for correcting our matrix element
and to acknowledge helpful discussions with Profs. J.~Blumlein
and J.~Kalinowski.

\bibliographystyle{prsty}
\bibliography{th-97-48}

\end{document}